\documentclass[epj]{svjour}
\usepackage{graphics}
\begin{document}
\bibliographystyle{unsrt}
%
\title{Strangeness Production in Heavy Ion Collisions at SPS Energies}

\author{Andr\'e Mischke \it{for the NA49 Collaboration}
\thanks{\email{a.mischke@phys.uu.nl}}
}
%
%
\institute{Gesellschaft f\"ur Schwerionenforschung, Darmstadt, Germany \\
present address: NIKHEF/University of Utrecht, The Netherlands}
\date{Received: \today / Revised version: date}
%
\abstract{
The NA49 collaboration has taken data of central Pb-Pb collisions at beam energies
from 20 to 158 GeV per nucleon. The large acceptance of the detector allows to study
particle yields in full phase space.
In this paper we present recent results on strangeness production 
(K, $\Lambda$ and $\phi$) in this energy range and compare to measurements
at lower and higher energies.
The K$^+/\pi^+$ and $\Lambda/\pi$ ratio shows a pronounced maximum around 
30~A$\cdot$GeV, whereas the K$^-/\pi^-$ and $\overline{\Lambda}/\pi$ ratio
exhibits a continuous rise.
The $\phi$/$\pi$ ratio also increases monotonically with beam energy. 
First results on multi-strange hyperon production at lower SPS energies are presented.
\PACS{
      {PACS-key}{discribing text of that key}   \and
      {PACS-key}{discribing text of that key}
     } 
}
\maketitle
%
\section{Introduction}
\label{intro}
Ultra-relativistic heavy ion collisions provide the environment to study nuclear
matter at high pressure and temperature. At energy densities of about 
1~GeV/fm$^3$ a phase transition of nuclear matter to a deconfined state of
strong\-ly interacting matter is predicted by lattice QCD calculations~\cite{Kar02}.
The measurement of strange baryons like $\Lambda$(uds) hyperons, which
contain between 30 and 60$\%$ of the total strangeness produced
(depending on the energy), allows to study simultaneously strangeness
production and the effect of baryon density in A-A collisions.
Essentially half of the $\bar{\rm s}$ quarks are contained in K$^+$.
The $\phi$ meson consists of a s$\bar{\rm s}$ pair (hidden strangeness),
and should therefore be more sensitive than kaons and lambdas to the 
production mechanism in the early stage of the collision.

\section{The NA49 experiment}
\label{sec:2}
Since 1994 the NA49 collaboration has investigated hadron production
in central Pb-Pb collisions at 158~A$\cdot$GeV. 
With\-in the framework of the NA49 energy scan programme~\cite{Add2}, 
started five years later, charged kaons, $\phi$ mesons as well as 
single and multi-strange hyperons were measured at lower energies
(20, 30, 40, 80~A$\cdot$GeV) over a large range of rapidity and 
transverse momentum. Detailed informations of the data sets can
be found in Ref.~\cite{Fri03}.

The NA49 detector is a large acceptance fixed target hadron 
spectrometer at the CERN-SPS~\cite{NIM99}. 
Tracking and particle identification by the measurement of the 
specific energy loss (d$E$/d$x$) is performed by two Time Projection 
Chambers (Vertex-TPC) located inside two vertex magnets 
(1.5 and 1.1~T, respectively) and two large volume TPCs 
situated downstream of the magnets symmetrically to the beam 
line. The relative d$E$/d$x$ resolution is 3-4~$\%$.
The dipole magnets (9 Tm bending power) allow the momentum 
determination with a resolution of
$\sigma(p)/p^2 = 0.3\cdot10^{-4}(\rm{GeV}/c)^{-1}$.
Two Time-of-Flight walls give additional particle identification near 
mid-rapidity ($\sigma_{t}$ = 60~ps).
The trigger on centrality is based on the measurement of the energy
deposited by the spectator nucleons in the forward calorimeter.

\section{Results}
\label{sec:2}
In the following, the results on kaon, $\Lambda$ hyperon and $\phi$
meson production in central Pb-Pb collisions at beam energies 
30-158~A$\cdot$GeV are presented. Starting from the reconstruction method
we will focus on the energy dependence of the particle multiplicities. 
For the results on transverse mass spectra and the extracted inverse slope 
parameter we refer to Refs.~\cite{Fri03,Mis03}.
\begin{figure*}
 \resizebox{0.9\textwidth}{!}{%
  \includegraphics{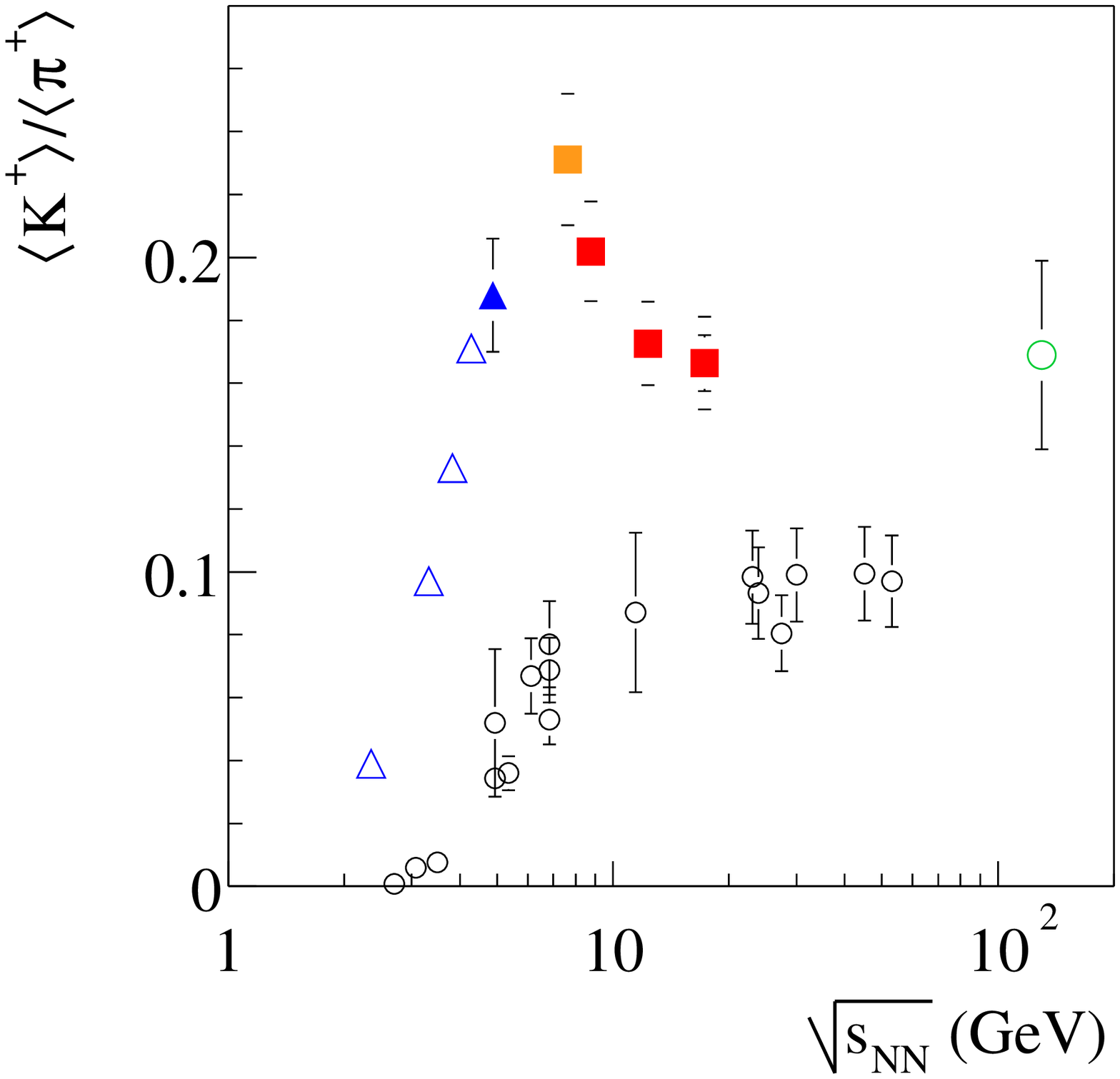}
  \hspace{6cm}
  \includegraphics{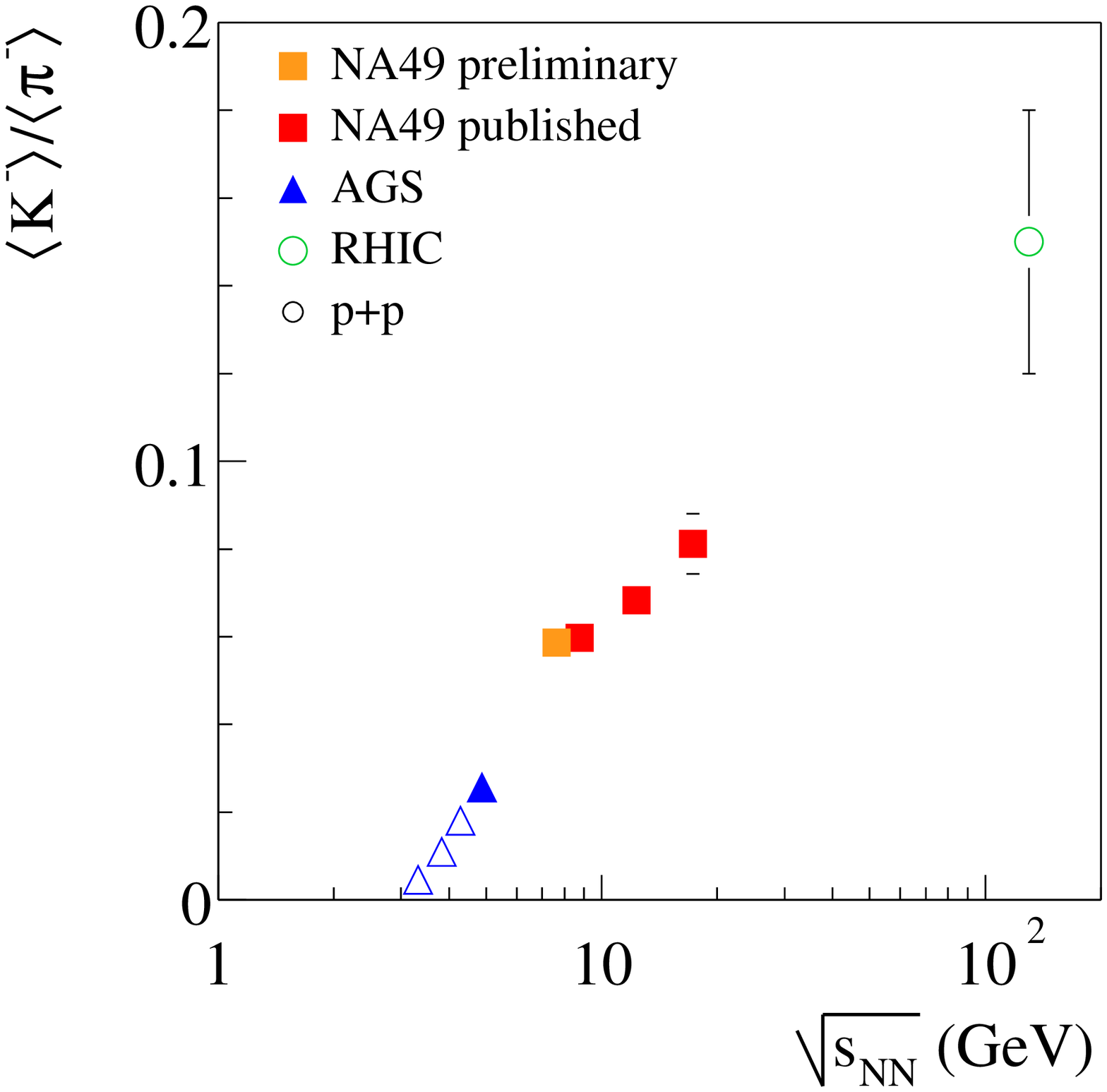}
 }
 \vspace{-0.35cm}
\caption{\protect \footnotesize 
The kaon-to-pion ratio as a function of cms energy in central Pb-Pb (Au-Au)
and p-p (open symbols) collisions. 
The AGS and RHIC measurements are taken from~\cite{KPI02}.}
\label{fig:1}
\end{figure*}
\begin{figure*}
 \resizebox{0.95\textwidth}{!}{%
  \includegraphics{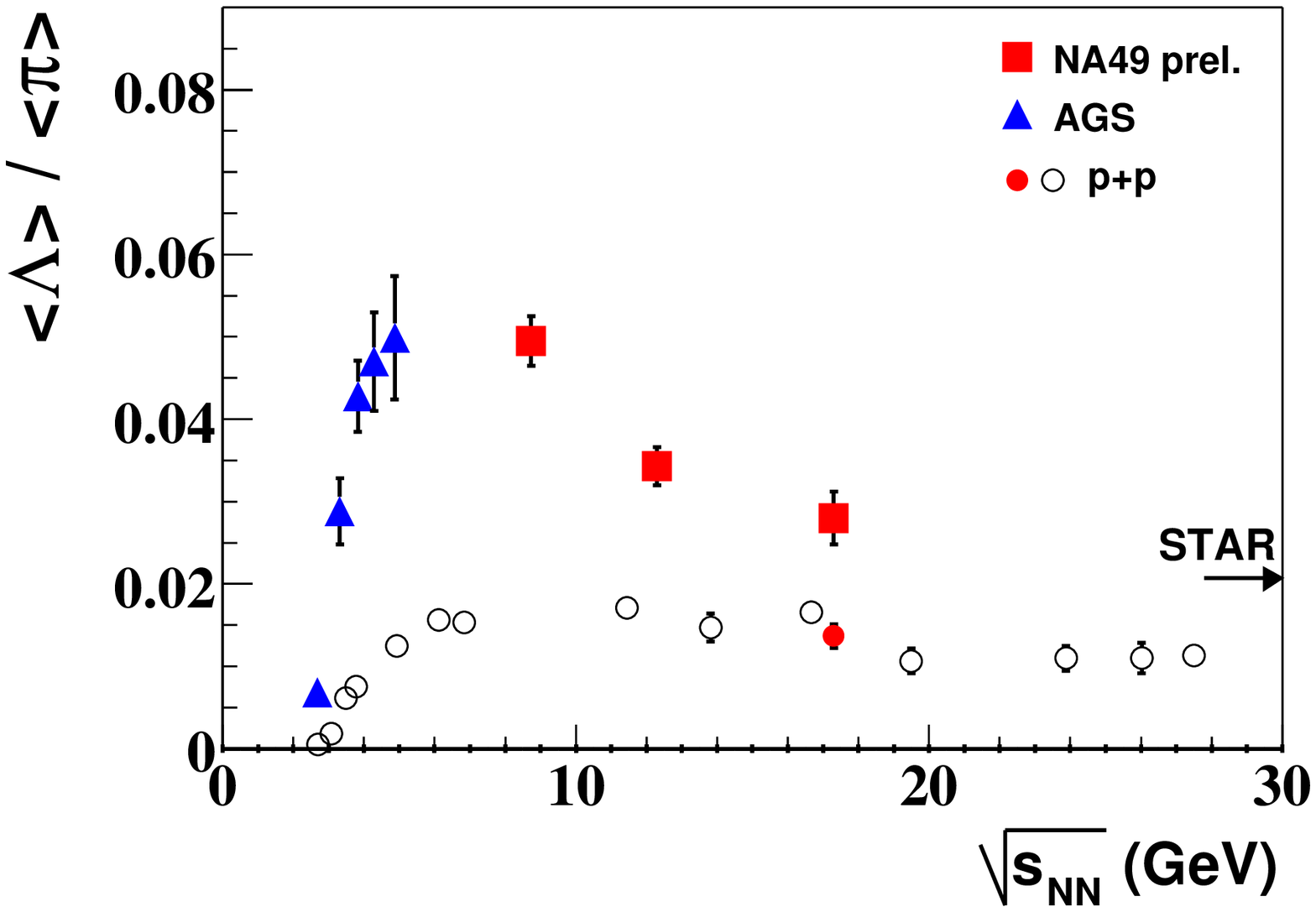}
  \includegraphics{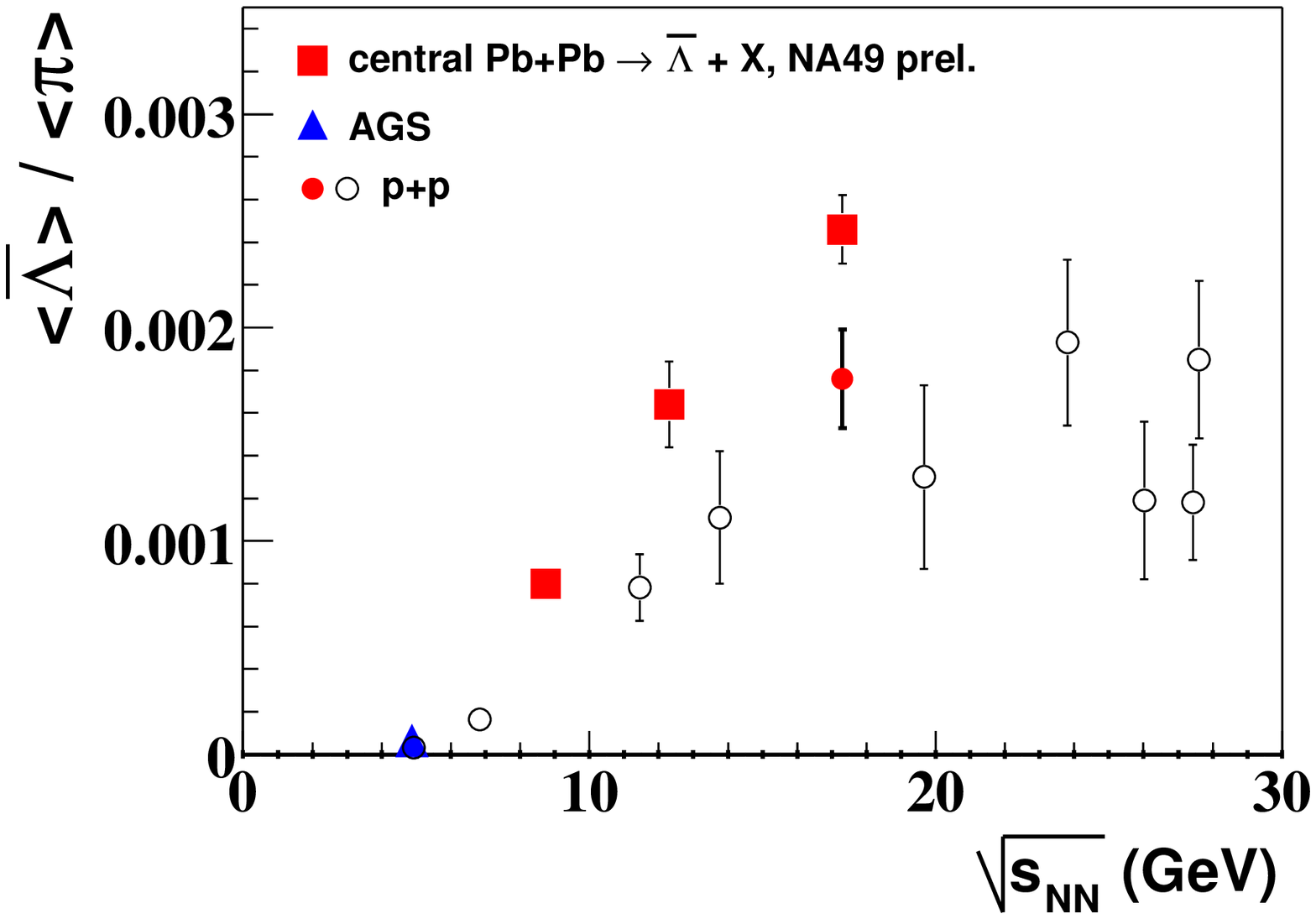}
 }
 \vspace{-0.35cm}
\caption{\protect \footnotesize 
Energy dependence of $\Lambda$-to-pion ratio in central Pb-Pb (Au-Au)
and p-p (open symbols) collisions. 
The AGS and RHIC measurements are taken from~\cite{Mis03}.
The STAR measurement at mid-rapidity is indicated by the arrow.}
\label{fig:2}
\end{figure*}

\subsection{Kaon production}
\label{sec:21}
Kaons are identified in NA49 by d$E$/d$x$ and ToF measurements depending
on their momentum. The raw data yields are corrected for geometrical
acceptance, in flight decay and efficiency of the detector system.
To compare the obtained multiplicities on kaon production at different 
energies and in elementary p-p collisions, respectively, the yields 
are normalized to the pion multiplicities. 
The $\pi^-$ multiplicities were extracted from the distribution of
all negatively charged hadrons by subtraction of the K$^-$, 
$\overline{\rm p}$ and e$^-$ contribution and secondary hadrons from
weak decays and interactions in the detector. This method is not 
applicable to the $\pi^+$ due to larger contributions from other 
positive particles. Therefore, the $\pi^+$ are calculated from the
$\pi^-$ yield assuming that the $\pi^+$/$\pi^-$ ratio, which is measured
in kinematical regions where both ToF and d$E$/d$x$ informations are
available, is constant in phase space (cf. Ref.~\cite{Fri03}).
The energy dependence of the K$^+/\pi^+$
and K$^-/\pi^-$ ratio is plotted in Fig.\ref{fig:1}. A sharp
maximum is observed at about 30~A$\cdot$GeV in the K$^+/\pi^+$ ratio,
whereas the K$^-/\pi^-$ exhibits a continuous rise with, perhaps,
a kink at the same energy.

\subsection{Lambda production}
\label{sec:22}
In NA49, neutral strange and multi-strange particles are identified 
by their decay topology 
$\Lambda \rightarrow p\pi^{-}$,
$\Xi \rightarrow \Lambda\pi$,
$\Omega \rightarrow \Lambda$K.
The charged decay products are measured with the TPCs.
$\Lambda$ hyperons from this analysis contain short-lived 
$\Sigma^{0}$, which decay electro-magnetically into $\Lambda\gamma$. 
The invariant mass distributions of $\Lambda$ and $\bar{\Lambda}$ 
hyperons are given in Ref.~\cite{Mis02}. The mass resolution 
($\sigma_{\rm m}$) of 2~MeV/$c^{2}$ is remarkably good. 
After corrections for acceptance and reconstruction efficiency the
$m_{\rm T}$ spectra and rapidity distributions were obtained
from the raw data yields (cf. Ref.~\cite{Mis03}). 
The total yields per event of the $\Lambda$ and $\overline{\Lambda}$ 
are obtained by integration of the distributions over rapidity and 
$p_{\rm T}$ with only small extrapolations into unmeasured regions.
The $\Lambda/\pi$ and $\overline{\Lambda}/\pi$ ratio as a function 
of collision energy $\sqrt{s_{\rm NN}}$ is shown in Fig.\ref{fig:2}, 
where $\pi = 1.5 (\pi^+ + \pi^-)$.
%
%
The $\Lambda/\pi$ ratio steeply increases at AGS energies, reaches 
a maximum and drops at SPS energies. 
Since ${\rm K}^{+}$ carry the major fraction of the produced
$\bar{\rm s}$ quarks one expect the ${\rm K}^{+}/\pi^{+}$ ratio to 
show a similar behavior as the $\Lambda/\pi$ ratio (using 
strangeness conservation) which is indeed the case (cf. Fig.\ref{fig:1}).

The enhancement of strangeness production in heavy ion collisions
compared to p-p is not a unique signature for the deconfined state,
since this enhancement is at low AGS energies, where a phase transition 
is not expected, seven times higher than at top SPS energies~\cite{MRTqm02}. 
Instead, rescattering processes like associated production 
$\pi$N $\rightarrow \Lambda$K play an important role at lower energies.

In comparison to the energy dependence of $\Lambda$ production, the 
$\overline{\Lambda}/\pi$ ratio shows a monotonic increase similar
to the K$^-/\pi^-$ ratio.
The measurements at 20 and 30~A$\cdot$GeV will clarify
whether there is also a structure like for the K$^-/\pi^-$ . 
The differences in the excitation function of $\Lambda$ and 
$\overline{\Lambda}$ can be attributed to their different production 
mechanisms and the effect of net-baryon density.                                  

\begin{figure}
 \resizebox{0.5\textwidth}{!}{%
  \includegraphics{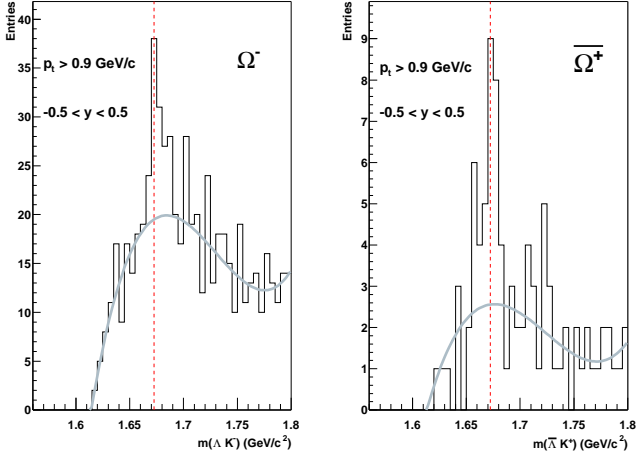}
 }
  \vspace{-0.3cm}
\caption{\protect \footnotesize 
The invariant mass distribution of $\Lambda$-K$^{-}$ 
(left) and $\bar{\Lambda}$-K$^{+}$ pairs (right) 
in central Pb-Pb collisions at 40~A$\cdot$GeV.}
\label{fig:3}
\end{figure}

First measurements of multi-strange hyperons at 40 A$\cdot$GeV beam 
energy have been shown (cf. Fig.\ref{fig:3})~\cite{Meu03,Mitr03}.
The available data sets from 20 to 158~A$\cdot$GeV allow to extract 
the $\Xi$ and $\Omega$ excitation function in the near future.

\subsection{$\phi$ meson production}
\label{sec:23}
\begin{figure}[t]
  \hspace{0.7cm}
 \resizebox{0.4\textwidth}{!}{%
  \includegraphics{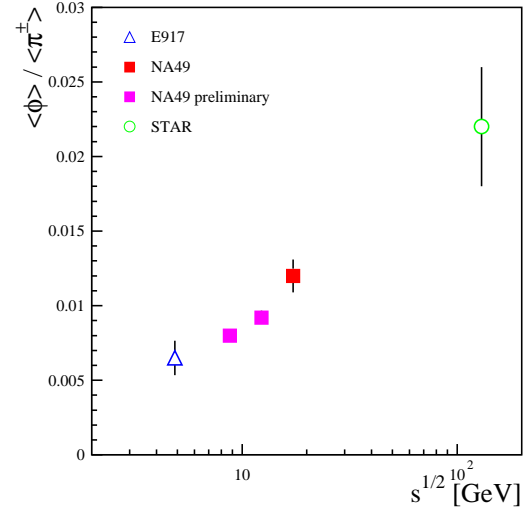}
}
  \vspace{-0.3cm}
\caption{\protect \footnotesize 
$\phi$ multiplicity ratios as a function of cms energy. 
The AGS and RHIC measurements are taken from~\cite{Fri03}.}
\label{fig:4}
\end{figure}

The $\phi$ meson is measured in NA49 via the invariant mass of its decay
products K$^+$K$^-$~\cite{Phi00}. The combinatorial background from 
random pairs is well described by means of the event-mixing method. 
The invariant mass distributions and the obtained rapidity spectra 
for 40, 80 and 158~A$\cdot$GeV are given in Refs.~\cite{Fri03,Phi00}. 
The extracted total $\phi$ yields normalized to the average number of pions 
$\pi^{\pm} = 0.5 (\pi^+ + \pi^-)$ are illustrated in Fig.\ref{fig:4}. 
This ratio shows the same monotonic rise from AGS to RHIC energies 
as the K$^-/\pi^-$ ratio.

\section{Summary and outlook}
\label{sec:3}
The excitation function of the K$^+/\pi^+$ and $\Lambda/\pi$ ratio in
central Pb-Pb (Au-Au) collisions shows a maximum around 30~A$\cdot$GeV 
whereas the K$^-/\pi^-$ and the $\overline{\Lambda}/\pi$ ratio exhibit 
a monotonic increase.

\begin{acknowledgement}
Acknowledgements: This work was supported by the Director, Office of
Energy Research,
Division of Nuclear Physics of the Office of High Energy and Nuclear
Physics of the US Department of Energy (DE-ACO3-76SFOOO98 and DE-FG02-91ER40609),
the Bundesministerium f\"ur Bildung und Forschung, Germany,
the Polish State Committee for Scientific Research (2 P03B 130 23,
SPB/CERN/P-03/Dz 446/2002-2004, 2 P03B 02418, 2 P03B 04123),
the Hungarian Scientific Research Foundation (T032648, T043514 and T32293),
Hungarian National Science Foundation, OTKA, (F034707),
and the Polish-German Foundation.
\end{acknowledgement}
%
%
%
\bibliography{../PhD/literatur}

\end{document}